\documentclass[11pt]{article}

\usepackage[utf8]{inputenc}
\usepackage{amsmath, amssymb, amsfonts, graphicx}
\usepackage[colorinlistoftodos]{todonotes}
\usepackage[colorlinks=true, allcolors=blue]{hyperref}
\usepackage{listings, color}
\usepackage{enumitem}
\usepackage{float}
\usepackage{textcomp}
\usepackage{authblk}
\usepackage[numbers,super]{natbib}
\usepackage{hyperref}
\usepackage{fullpage}

\begin{document}

\title{A regression algorithm for accelerated lattice QCD that exploits sparse inference on the D-Wave quantum annealer}

\author[1]{Nga T.T. Nguyen}
\author[1,2]{Garrett T. Kenyon}
\author[3\footnote{boram@lanl.gov}]{Boram Yoon}
\affil[1]{CCS-3, Information Sciences, Los Alamos National Laboratory, Los Alamos, NM 87545, USA}
\affil[2]{New Mexico Consortium, Los Alamos, NM 87545, USA}
\affil[3]{CCS-7, Applied Computer Science, Los Alamos National Laboratory, Los Alamos, NM 87545, USA}
\date{}

\maketitle
\begin{abstract}
We propose a regression algorithm that utilizes a learned dictionary optimized for sparse inference on a D-Wave quantum annealer. In this regression algorithm, we concatenate the independent and dependent variables as a combined vector, and encode the high-order correlations between them into a dictionary optimized for sparse reconstruction. On a test dataset, the dependent variable is initialized to its average value and then a sparse reconstruction of the combined vector is obtained in which the dependent variable is typically shifted closer to its true value, as in a standard inpainting or denoising task. Here, a quantum annealer, which can presumably exploit a fully entangled initial state to better explore the complex energy landscape, is used to solve the highly non-convex sparse coding optimization problem. The regression algorithm is demonstrated for a lattice quantum chromodynamics simulation data using a D-Wave 2000Q quantum annealer and good prediction performance is achieved. The regression test is performed using six different values for the number of fully connected logical qubits, between 20 and 64. The scaling results indicate that a larger number of qubits gives better prediction accuracy.

\end{abstract}

\flushbottom
\maketitle
\thispagestyle{empty}

\section{Introduction}\label{sec:intro}
Sparse coding refers to a class of unsupervised learning algorithms for finding an optimized set of basis vectors, or dictionary, for accurately reconstructing inputs drawn from any given dataset using the fewest number of non-zero coefficients. 
Sparse coding explains the self-organizing response properties of simple cells in the mammalian primary visual cortex~\cite{olshausen96,olshausen_sparse_1997}, and  has been successfully applied in various fields including image classification~\cite{yang2009,Coates:2011:IEV:3104482.3104598}, image compression~\cite{yijing}, and compressed sensing~\cite{Tao_Information,Donoho_Compressed}.  
Optimizing a dictionary $\boldsymbol{\phi}\in \mathbb{R}^{M\times N_q}$ for a given dataset and inferring optimal sparse representations $\boldsymbol{a}^{(k)}\in \mathbb{R}^{N_q}$ of input data  $\mathbf{X}^{(k)}\in \mathbb{R}^{M}$ involves finding solutions of the following minimization problem:
\begin{align}
	\min\limits_{ \boldsymbol{\phi} }\sum_{k}\min\limits_{ \boldsymbol{a}^{(k)} } \left[  \, \frac{1}{2}  ||  \mathbf{X}^{(k)} - \boldsymbol{\phi} \boldsymbol{a}^{(k)} ||^2	+	\lambda || \boldsymbol{a}^{(k)} ||_0 \, \right]\,,
	\label{eq:H_SC}
\end{align}
where $k$ is the index of the input data, and $\lambda$ is the sparsity penalty parameter. Note that the convergence of the solution is guaranteed only when the norm of column vectors of the dictionary $\boldsymbol{\phi}$ is constrained by an upper bound, which is unity in this study.
Because of the $L_0$-norm, the minimization problem falls into an NP-hard complexity class with multiple local minima~\cite{Natarajan:1995:SAS:207985.207987} in the energy landscape.

Recently, we developed a mapping of the \mbox{$\boldsymbol{a}^{(k)}$-optimization} in Eq.~\eqref{eq:H_SC} to the quadratic unconstrained binary optimization (QUBO) problem that can be solved on a quantum annealer and demonstrated its feasibility on the D-Wave systems~\cite{Nguyen:2016,8123653,Nguyen2018ImageCU}. The quantum processing unit of the D-Wave systems realizes the quantum Ising spin system in a transverse field and finds the lowest or the near-lowest energy states of the classical Ising model,
\begin{equation}
	H(\boldsymbol{h}, \boldsymbol{J}, \boldsymbol{s}) = \sum_i^{N_q} {h_i s_i} + \sum_{i<j}^{N_q} {J_{ij}  s_i s_j }\,,
	\label{eq:H_Dwave}
\end{equation}
using quantum annealing~\cite{PhysRevE.58.5355,Finnila_1994,DWAVE}. Here $s_{i}=\pm1$ is the binary spin variable, $h_i$ and $J_{ij}$ are the qubit biases and coupling strengths that can be controlled by a user, and optimization for the Ising model is isomorphic to a QUBO problem with $a_i = (s_i+1)/2$. By mapping the sparse coding to a QUBO structure, the sparse coefficients are restricted to binary variables $a_i\in\{0,1\}$, and it makes the $L_0$-norm equivalent to the $L_1$-norm. Despite this restriction, it was able to provide good sparse representation for the the MNIST~\cite{lecun-mnisthandwrittendigit-2010,Nguyen:2016,Nguyen2018ImageCU} and \mbox{CIFAR-10}~\cite{cifar10,8123653} images.

In this paper, we propose a regression algorithm using the sparse coding on D-Wave 2000Q in Section~\ref{sec:reg} and apply the algorithm to a prediction of quantum chromodynamics (QCD) simulation observable in Section~\ref{sec:exp}.

\section{\label{sec:reg} Regression algorithm using sparse coding on D-Wave 2000Q}
\subsection{\label{subsec:reg}Regression model}
Consider $N$ sets of training data $\{\mathbf{X}^{(i)}, y^{(i)}\}_{i=1}^N$, and $M$ sets of the test data $\{\mathbf{X}^{(j)}\}_{j=1}^M$, where $\mathbf{X}^{(i)}\equiv\{x_1^{(i)}, x_2^{(i)}, \ldots, x_D^{(i)}\}$ is an input vector known as the \emph{independent variable}, and $y^{(i)}$ is an output variable known as the \emph{dependent variable}. A regression model $F$ can be built by learning correlations between the input and output variables on the training dataset, so that it can make predictions $\hat{y}$ of $y$ for an unseen input data $\mathbf{X}$ as
\begin{align}
  F(\mathbf{X}) = \hat{y} \approx y\,.
\end{align}
Such a regression model can be built using the sparse coding learning implemented on a quantum annealer described below.

\begin{itemize}
\item \textbf{Pre-training}
\begin{itemize}[leftmargin=1em]
  \item[(1)] Normalize $x_d^{(i)}$ and $y^{(i)}$ so that their standard deviations become comparable. One possible choice is rescaling the data to have a zero mean and a unit variance using the sample mean and sample variance of the training dataset. This procedure is an essential step for the regression algorithm as it makes the reconstruction error for each component comparable.
  \item[(2)] Using $\mathbf{X}$ in the test dataset ($M$) or those in the combined training and test datasets ($N+M$), perform sparse coding training and obtain the dictionary $\boldsymbol{\phi}$ for $\mathbf{X}$.
\end{itemize}
\item \textbf{Training}
\begin{itemize}[leftmargin=1em]
\sloppy
  \item[(3)] Concatenate the input and output variables of the training dataset and build the concatenated vectors 
  $\widetilde{\mathbf{X}}^{(i)}\equiv\{x_1^{(i)}, x_2^{(i)}, \ldots, x_D^{(i)}, y^{(i)}\}$. 
  Extend the dictionary matrix $\boldsymbol{\phi} \in \mathbb{R}^{D\times N_q}$ obtained in the pre-training to  $\widetilde{\boldsymbol{\phi}}_o \in \mathbb{R}^{(D+1)\times N_q}$, filling up the new elements by zeros.
  \item[(4)] Taking $\widetilde{\mathbf{X}}^{(i)}$ as the input signal and $\widetilde{\boldsymbol{\phi}}_o$ as an initial guess of the dictionary, perform sparse coding training on the training dataset and obtain the dictionary $\widetilde{\boldsymbol{\phi}}$. Through this procedure, $\widetilde{\boldsymbol{\phi}}$ will encode the correlation between $x_d^{(i)}$ and $y^{(i)}$.
\end{itemize}
\item \textbf{Prediction}
\begin{itemize}[leftmargin=1em]
\item[(5)] For the test dataset, for which only $\mathbf{X}^{(j)}$ is given, build a vector
  $\widetilde{\mathbf{X}}_o^{(j)}\equiv\{x_1^{(j)}, x_2^{(j)}, \ldots, x_D^{(j)}, \bar{y}^{(j)}\}$,
  where $\bar{y}^{(j)}$ is an initial guess of $y^{(j)}$. One possible choice of $\bar{y}^{(j)}$ is the average value of $y^{(i)}$ in the training dataset.
  \item[(6)] Using the dictionary $\widetilde{\boldsymbol{\phi}}$ obtained in (4), find a sparse representation $\boldsymbol{a}^{(j)}$ for $\widetilde{\mathbf{X}}^{(j)}_o$ and calculate reconstruction as $\widetilde{\mathbf{X}}'^{(j)} = \widetilde{\boldsymbol{\phi}}\boldsymbol{a}^{(j)}$. This replaces the outlier components, including $\bar{y}^{(j)}$, in $\widetilde{\mathbf{X}}^{(j)}_o$ by the values that can be described by $\widetilde{\boldsymbol{\phi}}$.
  \item[(7)] After inverse-normalization, the $(D+1)$'th component of $\widetilde{\mathbf{X}}'^{(j)}$ is the prediction of $y^{(j)}$: ${(\widetilde{\mathbf{X}}'^{(j)})_{D+1} = \hat{y}^{(j)} \approx y^{(j)}}$.
\end{itemize}
\end{itemize}

In this regression model, $D$ should be sufficiently large so that the initial guess of the dependent variable $\bar{y}_j$ does not bias the reconstruction. This procedure can be extended to predict multiple variables by increasing the dimension of $y$, in exchange for prediction accuracy.

\subsection{Sparse coding on a D-Wave quantum annealer}
The \mbox{$\boldsymbol{a}^{(k)}$-optimization} of the sparse coding problem in Eq.~\eqref{eq:H_SC}, can be mapped onto the D-Wave problem in Eq.~\eqref{eq:H_Dwave}, by the following transformations~\cite{Nguyen:2016,8123653,Nguyen2018ImageCU}: 
\begin{equation}
	h_i = (-\boldsymbol{\phi}^{T}  \mathbf{X} + (\lambda + \frac{1}{2}))_i\,, \qquad
	J_{ij} 	= (\boldsymbol{\phi}^{T} \boldsymbol{\phi})_{ij}\,, \qquad
	s_i = 2a_i-1\,.
\label{eq:hQ}
\end{equation}
In this mapping, each neuron, the sparse coefficient, of the sparse coding model corresponds to a qubit. After a measurement, the quantum state of a qubit collapses to $0$ or $1$, which indicates that the neuron can have only two states of fire ($1$) or silent ($0$).
Here the qubit-qubit coupling $\boldsymbol J$ shares similarity with the lateral neuron-neuron inhibition in the locally competitive algorithm~\cite{rozell.06c}, and the constant $\lambda$ makes the solution sparse by acting a constant field forcing the qubits to stay in $a_i=0$ ($s_i=-1$) state. By performing the quantum annealing for a given dictionary $\boldsymbol{\phi}$ and input data vector $\mathbf{X}$ with the transformations given in Eq.~\eqref{eq:hQ}, one can obtain the optimal sparse representation $\boldsymbol{a}$.

\begin{figure}
  \begin{center}
  	\includegraphics[width=0.39\textwidth]{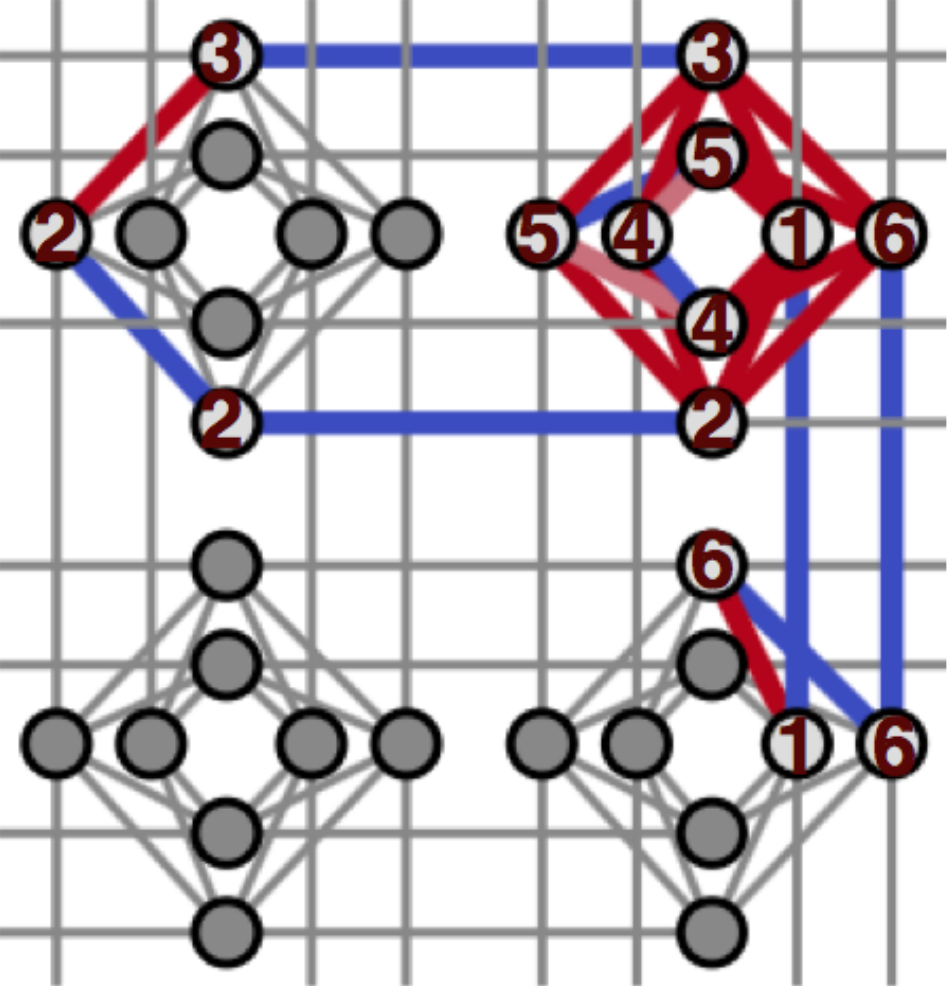}
  \end{center}
  \caption{A subset ($1/64$) of the Chimera structure of the D-Wave 2000Q consisting of $32$ qubits (circles) arranged in a $2\times2$ matrix of unit cells of 8 qubits. The qubits within a unit cell have relatively dense connections, while the interactions between the unit cells can be made through the sparse connections in their edges. This figure also shows an example of embedding $6$ fully-connected logical qubits (numbers from $1$ to $6$ inside $14$ circles) onto the D-Wave chimera using $14$ physical qubits, in which red edges indicate bipartite couplings between qubits while blue edges indicate chained qubits. After such embedding, for example, the logical qubit $1$ is mapped to two physical qubits tiled from one qubit in the top right and one in the bottom right unit cell, while the logical qubit $2$ mapped to three physical qubits tiled from two qubits in the top left and one qubit in the top right unit cell, and so forth. } 
  \label{fig:embed_6qubits}
\end{figure}

An ideal D-Wave 2000Q consists of 2048 qubits, and the entire coupling graph of this 2048-qubit system is called the perfect Chimera 2000Q, whose 1/16 subset is illustrated in Fig.~\ref{fig:embed_6qubits}. However, the graph is sparsely-connected in which one qubit can couple to only up to 6 other qubits. With the limited connectivity between the qubits, a perfect 2048-qubit Chimera has 6016 couplers. To map a general Ising model problem with arbitrary bipartite couplings to a D-Wave Chimera, in many cases, one requires an additional step called the \textit{embedding}. In the case of the sparse coding problem,  the embedding translates a graph of a fully-connected logical qubits to the Chimera graph of the partially-connected physical qubits by chaining a group of physical qubits together with a certain chain strength $\xi$. One example of such a mapping of fully-connected logical 6 qubits to the D-Wave 2000Q by chaining 14 physical qubits is described in Figure~\ref{fig:embed_6qubits}. This  embedding procedure results in a significant reduction of the total available logical qubits that represent the mapped problem; on a perfect D-Wave 2000Q QPU, only up to 65 fully-connected logical qubits can be mapped. In practice, however, some qubits on the QPU are inoperable after a calibration, and the maximum number of logical qubits that could be embedded decreases. For example, the D-Wave 2000Q quantum annealer at Los Alamos National Laboratory (LANL) has only 2032 active qubits with 5924 active couplers. We find that an arbitrary QUBO problem up to 64 fully-connected logical qubits can be embedded in the LANL D-Wave 2000Q.

%


\section{Application to lattice QCD\label{sec:exp}}
QCD is a theory of quarks and gluons, which are the fundamental particles composing hadrons such as pions and protons, and their interactions. It is a part of the Standard Model of particle physics, and the theory has been demonstrated by a large class of experiments over the decades \cite{Patrignani:2016xqp,Greensite:2011zz}. Lattice QCD is a discrete formulation of QCD on a Euclidean space time lattice, which allows us to solve low-energy QCD problems using computer simulations by carrying out the Feynman path integration using Monte Carlo methods \cite{Wilson:1974sk,Creutz:1980zw}.

In lattice QCD simulations, a large number of observables are calculated over an ensemble of the Gibbs samples of gluon fields, called the lattices, and computational cost for calculating those observables is expensive in modern simulations. However, the observables' fluctuations over the statistical samples of the lattices are correlated as they share the same background lattice. By exploiting the correlation between them, in Ref.~\cite{Yoon:2018krb}, Gradient Tree Boosting (GTB) regression algorithm was able to replace the computationally expensive direct calculation of some observables by the computationally cheap machine learning predictions of them from other observables.

In this section, we apply the regression algorithm proposed in Section~\ref{sec:reg} to the lattice QCD simulation data used for the calculation of the charge-parity (CP) symmetry violating phase $\alpha_{\textrm{CPV}}$ of the neutron~\cite{Yoon:2017tag,Pospelov:2005pr}. Here we consider three types of observables: (1) two-point correlation functions of neutrons calculated without CP violating (CPV) interactions $C_{\textrm{2pt}}$, (2) $\gamma_5$-projected two-point correlation functions of neutrons calculated without CPV interactions $C_{\textrm{2pt}}^P$, and (3) $\gamma_5$-projected two-point correlation functions of neutrons calculated with CPV interactions $C_{\textrm{2pt}}^{P,\textrm{CPV}}$, and the phase $\alpha_{\textrm{CPV}}$ is extracted from the imaginary part of $C_{\textrm{2pt}}^{P,\textrm{CPV}}$. Those observables are calculated at multiple values of the nucleon source and sink separations in Euclidean time direction $t$. 

\subsection{Method}

Our goal of the regression problem is to predict the imaginary part of $C_{\textrm{2pt}}^{P,\textrm{CPV}}$ at $t=10a$ from the real and imaginary parts of the two-point correlation functions calculated without CPV interactions, $C_{\textrm{2pt}}$ and $C_{\textrm{2pt}}^P$, at $t=8a, 9a, 10a, 11a,$ and $12a$, where $a$ is the lattice spacing. It forms a problem with single value of output variable ($y$) and 20 values (two observables, real/imag, 5 timeslices) of the input variables ($\mathbf{X}$). In this application, we use 15616 data points of of these observables measured in Refs.~\cite{Bhattacharya:2016oqm,Bhattacharya:2016rrc} divided into 6976 training data and 8640 test data.  Using these datasets, we follow the regression procedure proposed in Section~\ref{sec:reg} to predict $y$ of the test dataset that contains around 9K data points.

The procedure can be summarized as follows. First, we standardize the total data using the mean and variance of the training dataset for normalization. Then, we perform the pre-training and obtained $\boldsymbol{\phi}$ for the 20 elements of $\mathbf{X}$ only using the test dataset. After appending the $y$ to $\mathbf{X}$ as the 21st element in the training dataset, we perform the sparse coding dictionary learning and update $\boldsymbol{\phi}$ to encode correlation between $\mathbf{X}$ and $y$. For prediction, input vectors $\mathbf{X}$ in the test dataset are augmented to dimension of 21 vectors by appending the average value of $y$, which is 0 after standardization. Finally, sparse coefficients $\boldsymbol{a}$ for the augmented input vectors are calculated with the fixed dictionary $\boldsymbol{\phi}$ obtained above, and predictions of $y$ are estimated by taking the 21st element of the reconstructed vectors on the test dataset.
\begin{figure}[tb]
  \vspace{-0.0in}
  \hspace{0.in}
  \begin{center}
  	\includegraphics[trim=0 255 0 0, clip, width=0.45\textwidth]{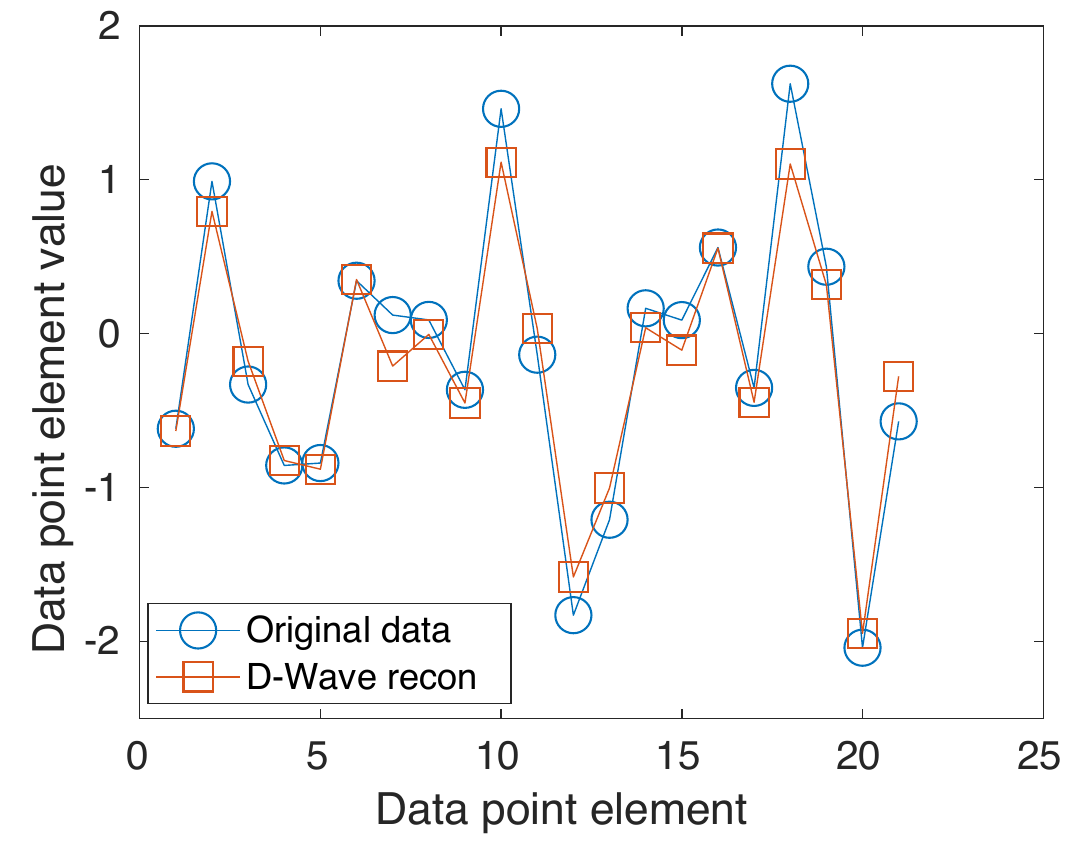}
  	\qquad
  	\includegraphics[trim=0 0 0 255, clip, width=0.45\textwidth]{Figures/Image134_and_Image2224_predict_replot.pdf}
  \end{center}
  \vspace{-0.in}
  \caption{Original, or ground truth, data (blue circles) and the reconstruction from the missing-21st-element data using \mbox{D-Wave} 2000Q with $N_q = 64$ (red squares) for two randomly chosen data points. Here the $21$st-element is the dependent variable of the prediction, whose initial value before the reconstruction is given by 0.  }
  \vspace{-0.2in}
  \label{fig:imrecon}
\end{figure}

Note that a sparse coding problem solves for the sparsest representation $\boldsymbol{a}$ and the dictionary $\boldsymbol{\phi}$, simultaneously, by minimizing Eq.~\eqref{eq:H_SC}.  
First, our optimization for $\boldsymbol{a}$ is performed using the D-Wave 2000Q at a given $\boldsymbol \phi$, whose initial guess is given, in general, by random numbers or via imprinting technique.  Then, the optimization for $\boldsymbol{\phi}$ is performed on classical CPUs.  The latter step is an offline learning for the fixed values of $\boldsymbol a$ obtained using D-Wave 2000Q. In the offline learning procedure, $\boldsymbol{\phi}$ is learned using the batch stochastic gradient descent (SGD) algorithm:
\begin{equation}
    \boldsymbol \phi : = \boldsymbol \phi - \eta \frac{\partial E_b}{\partial \boldsymbol \phi}\,,
    \label{eq:SGD}
\end{equation}
where $E_b = \frac{1}{n_b}\sum_{i=1}^{n_b}{E_i}$ with $E_i$ is the sparse coding energy function for a given input data given in Eq.~\eqref{eq:H_SC}, and $\eta$ is the learning rate. In this study, $\eta$ is initially set to 0.01 and gradually decreased during the training procedure.  Batch-learning is used with the batch size of $n_b = 50$. We repeat the iterative update of the quantum D-Wave inference for $\boldsymbol{a}$ and SGD learning for $\boldsymbol{\phi}$ until a convergence is attained. On average, we find the convergence after $4$ or $5$ iterations. In this study, we use the SAPI2 python client libraries~\cite{sapi} for implementing D-Wave operations.

The sparsity of the sparse representation $\boldsymbol{a}$ associated with the sparsity penalty parameter $\lambda$ is calculated by the ratio of nonzero elements in $\boldsymbol{a}$. In this study, $\lambda$ is tuned to the values that make the average sparsity about 20\%, because we find that the 20\% of sparsity provides an optimal prediction performance, after examining a few different values of $\lambda$. This corresponds to $\lambda$ = [0.06, 0.1], which we varied for different $N_q$ studies in our experiments.  Although the prediction performance could be further optimized by an extensive parameter search, such as that performed in Ref.~\cite{Kenyon_phasetransition}, the procedure is computationally expensive so ignored in this proof-of-principle study.

Note that the definition of the overcompleteness is not straightforward for the D-Wave inferred sparse coding because the input signal $\mathbf{X}$ may have arbitrary real numbers, while the sparse coefficients $\boldsymbol{a}$ could have only binary numbers of 0 or 1. Ignoring the subtlety, for simplicity, the overcompleteness $\gamma$ for the input signal of dimension 20 (or 21 for extended vectors) can be calculated by $\gamma = N_q/20$.

\subsection{Results}
Examples of the reconstruction and prediction from the randomly chosen test data points are visualized in Figure~\ref{fig:imrecon}. In the plot, the first 20 elements are the input variables, and the element 21 is the output of the prediction algorithm. As one can see, the reconstruction of the $21$st element, which was 0 in their initial guess, is successfully shifted close to their ground truth, as expected.

\begin{figure*}[tb]
  \begin{center}
  	\includegraphics[width=0.99\textwidth]{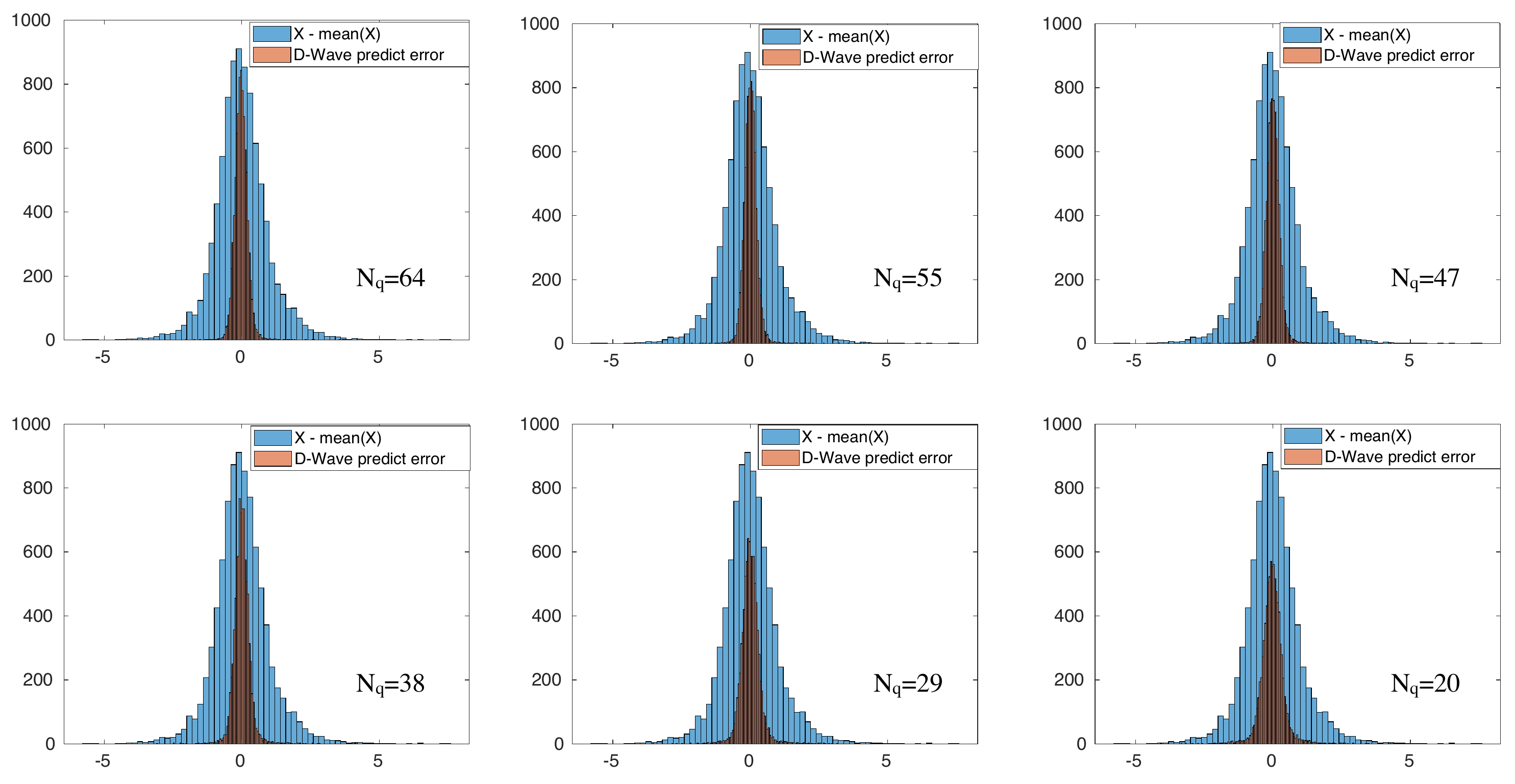}
  \end{center}
  \caption{Distribution of the prediction error $\Delta^{(i)}$ of the 21st element plotted against the distribution of the ground truth for different numbers of qubits $N_q=20, 29, 38, 47, 55$, and $64$.  The narrower width of the prediction error indicates the better prediction.  Standard deviations of the prediction errors for $N_q= 20, 29, 38, 47, 55$, and $64$ are $0.41$, $0.375$, $0.319$, $0.29$, $0.273$ and $0.254$, respectively. Scaling of the prediction error is summarized in Figure~\ref{fig:std_Nq}. }
  \label{fig:std_Nq_plot}
\end{figure*}
In order to investigate the prediction accuracy for different $N_q$, we explore the prediction algorithm with six different numbers of qubits $N_q=20, 29, 38, 47, 55$ and $64$, which corresponds to $\gamma \approx 1 \sim 3$. Note that the larger $N_q$ implies the more difficult optimization problem, and $N_q=64$ is the maximum number of logical qubits that can be embedded onto the D-Wave 2000Q. In the experiments with the D-Wave 2000Q, we use annealing time $\tau = 20 \mu s$, which is a relatively short annealing time. In addition, we run the experiments with 10 different values of the chain strengths $\xi$ for each input data point to obtain an optimal solution in exchange for longer wallclock time. We performed our D-Wave experiments using 20 reads for each value of $\xi$ and take the lowest energy solution. In Figure~\ref{fig:std_Nq_plot}, we show the distribution of the normalized original data of the dependent variable $y^{(i)}$ and its prediction error  $\Delta^{(i)}$ defined by the difference between the ground truth $y^{(i)}$ and its prediction $\hat{y}^{(i)}$: ${\Delta^{(i)} = y^{(i)} - \hat{y}^{(i)}}$. It is clearly demonstrated that (1) the prediction error is much smaller than the fluctuation of the original data, (2) the prediction error is sharply distributed near 0, which indicates no obvious bias in the prediction, and (3) the prediction error tends to be smaller when $N_q$ becomes larger.

To evaluate the prediction quality, the recovery of the $21$st element in the extended input vector, quantitatively, we calculate the ratio of the standard deviations of the prediction error and that of the original data: ${Q \equiv \sigma(\Delta) / \sigma(y)}$.  $Q$ converges to $0$ when the prediction is precise, and $Q\ge 1$ indicates no prediction for a statistical data. Note that this definition of the prediction quality does not account for the bias of the prediction because the bias for the prediction of a statistical data can be removed by following the procedure introduced in Ref.~\cite{Yoon:2018krb} based on the variance reduction technique for lattice QCD calculations~\cite{Bali:2009hu,Blum:2012uh}.

\begin{figure}[tb]
\centering
  	\includegraphics[width=0.45\textwidth]{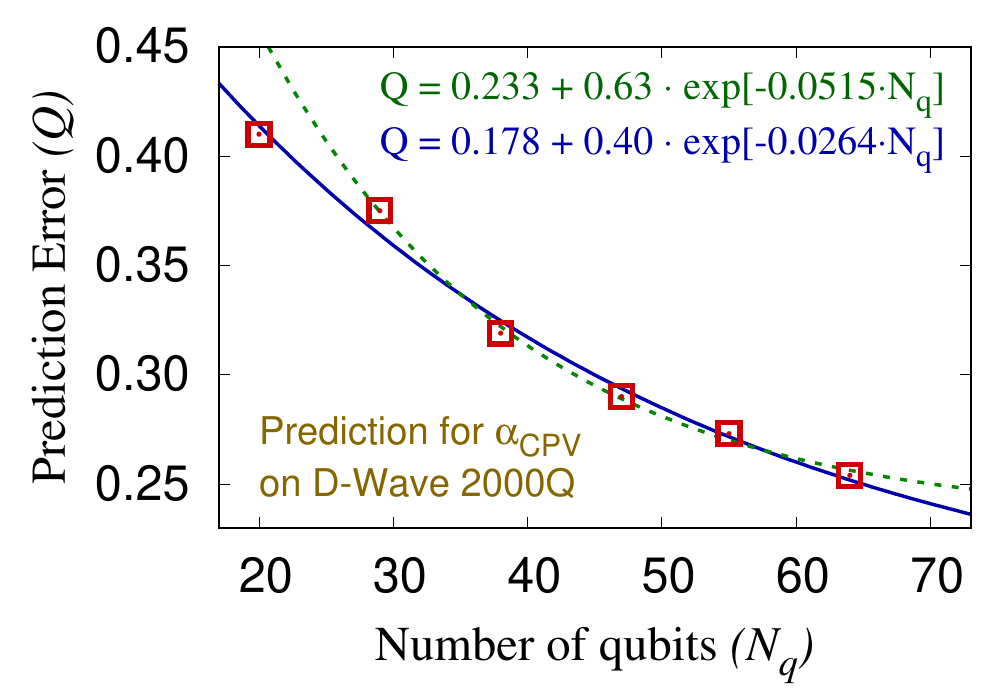}
    \caption{Prediction error $Q$ with the sparse coding regression algorithm implemented on D-Wave 2000Q applied to prediction of the lattice QCD simulation data as a function of $N_q$ (red squares). An exponential ansatz is fitted to the data points for all $N_q$ (blue solid line) and those excluding $N_q=20$ (green dashed line).
  }
  \label{fig:std_Nq}
\end{figure}

Figure~\ref{fig:std_Nq} shows the prediction error $Q$ as a function of the number of qubits.  It is clearly demonstrated that the systematic decrease of the prediction error as $N_q$ is increased. Although no theory explaining the scaling is known, we find that the scaling roughly follows the exponential decay ansatz $Q_\infty+B\cdot\exp[-C\cdot N_q]$. By fitting the ansatz to the data points, an asymptotic value of the prediction quality is obtained as $Q_\infty \approx 0.18$ or $0.23$ for $N_q \to \infty$, depending on whether we include $N_q=20$ data point or not in the fit. For a comparison, regression algorithms provided by the scikit-learn library~\cite{scikit-learn} on a classical computer are investigated for the same dataset, and GTB regression algorithm~\cite{breiman1984classification,Friedman00greedyfunction,Friedman:2002:SGB:635939.635941} showed the best prediction performance with $Q=0.15(1)$.

The data points in Figure~\ref{fig:std_Nq} are obtained with a fixed training and test datasets without cross validation because of the limited D-Wave 2000Q resources. However, the classical regression algorithms we applied on the same datasets showed prediction quality $Q$ between 0.15 -- 0.33 with 2\% -- 8.6\% uncertainties, where the uncertainties are estimated using the bootstrap resampling method following Ref.~\cite{Yoon:2018krb}. Based on this observation, we expect smaller than 10\% uncertainties for the data points presented in Figure~\ref{fig:std_Nq}. 

Pre-training is demonstrated to lower the prediction error of this regression algorithm, significantly. When performed the prediction with $N_q=64$ qubits without the pre-training procedure, we find that $Q=0.34$, while it becomes $Q=0.254$ with the pre-training. Without the pre-training, furthermore, we find that the required number of iterative updates of the D-Wave inference for $\boldsymbol{a}$ and SGD learning for $\boldsymbol{\phi}$ is increased to about 10 iterations.

\section{Conclusion}
In this paper, we proposed a regression algorithm using sparse coding dictionary learning that can be implemented on a quantum annealer, based on the formulation of a regression as an inpainting problem. A pre-training technique is introduced to improve the prediction quality. The procedure is described in Section~\ref{subsec:reg}. The regression algorithm was numerically demonstrated using a set of lattice QCD simulation observables and was able to predict the correlation function calculated in the presence of the CPV interactions from those calculated without the CPV interaction. The regression experiment is carried out using the D-Wave 2000Q quantum annealer with minor embedding technique in order to obtain fully-connected logical qubits. The study is performed for six different values of the number of qubits between 20 and 64, and it showed a systematic decrease of the prediction error as the number of qubits is increased (see Figure~\ref{fig:std_Nq}). With a larger number of qubits and elaborately tuned the sparsity parameter, we expect further improved performance in future.
\vfill

\bibliographystyle{apsrev4-1}
\bibliography{ref}

\section*{Acknowledgments}
The sparse coding optimizations were carried out using the D-Wave 2000Q at Los Alamos National Laboratory (LANL). Simulation data used for the numerical experiment were generated using the computer facilities at (i)the National Energy Research Scientific Computing Center, a DOE Office of Science User Facility supported by the Office of Science of the U.S. Department of Energy under Contract No. DE-AC02-05CH11231; and, (ii) the Oak Ridge Leadership Computing Facility at the Oak Ridge National Laboratory, which is supported by the Office of Science of the U.S. Department of Energy under Contract No. DE-AC05-00OR22725; (iii) the USQCD Collaboration, which is funded by the Office of Science of the U.S. Department of Energy, (iv) Institutional Computing at Los Alamos National Laboratory. This work was supported by the U.S. Department of Energy, Office of Science, Office of High Energy Physics under Contract No. 89233218CNA000001. BY also acknowledges support from the U.S. Department of Energy, Office of Science, Office of Advanced Scientific Computing Research and Office of Nuclear Physics, Scientific Discovery through Advanced Computing (SciDAC) program, and the LANL LDRD program.

\end{document}